\documentclass[prd,
               twocolumn,
               eqsecnum,
               showpacs,
               letterpaper,
               superscriptaddress,
               altaffilletter]
               {revtex4}

\usepackage{graphicx,amssymb,amsmath,times,latexsym}

\begin{document}

\title{Search for High Frequency Gravitational Wave Bursts in the First Calendar Year of LIGO's Fifth Science Run}


\newcommand*{\AG}{Albert-Einstein-Institut, Max-Planck-Institut f\"{u}r Gravitationsphysik, D-14476 Golm, Germany}
\affiliation{\AG}
\newcommand*{\AH}{Albert-Einstein-Institut, Max-Planck-Institut f\"{u}r Gravitationsphysik, D-30167 Hannover, Germany}
\affiliation{\AH}
\newcommand*{\AU}{Andrews University, Berrien Springs, MI 49104 USA}
\affiliation{\AU}
\newcommand*{\AN}{Australian National University, Canberra, 0200, Australia}
\affiliation{\AN}
\newcommand*{\CH}{California Institute of Technology, Pasadena, CA  91125, USA}
\affiliation{\CH}
\newcommand*{\CA}{Caltech-CaRT, Pasadena, CA  91125, USA}
\affiliation{\CA}
\newcommand*{\CU}{Cardiff University, Cardiff, CF24 3AA, United Kingdom}
\affiliation{\CU}
\newcommand*{\CL}{Carleton College, Northfield, MN  55057, USA}
\affiliation{\CL}
\newcommand*{\CS}{Charles Sturt University, Wagga Wagga, NSW 2678, Australia}
\affiliation{\CS}
\newcommand*{\CO}{Columbia University, New York, NY  10027, USA}
\affiliation{\CO}
\newcommand*{\ER}{Embry-Riddle Aeronautical University, Prescott, AZ   86301 USA}
\affiliation{\ER}
\newcommand*{\EU}{E\"{o}tv\"{o}s University, ELTE 1053 Budapest, Hungary}
\affiliation{\EU}
\newcommand*{\HC}{Hobart and William Smith Colleges, Geneva, NY  14456, USA}
\affiliation{\HC}
\newcommand*{\IA}{Institute of Applied Physics, Nizhny Novgorod, 603950, Russia}
\affiliation{\IA}
\newcommand*{\IU}{Inter-University Centre for Astronomy  and Astrophysics, Pune - 411007, India}
\affiliation{\IU}
\newcommand*{\HU}{Leibniz Universit\"{a}t Hannover, D-30167 Hannover, Germany}
\affiliation{\HU}
\newcommand*{\CT}{LIGO - California Institute of Technology, Pasadena, CA  91125, USA}
\affiliation{\CT}
\newcommand*{\LO}{LIGO - Hanford Observatory, Richland, WA  99352, USA}
\affiliation{\LO}
\newcommand*{\LV}{LIGO - Livingston Observatory, Livingston, LA  70754, USA}
\affiliation{\LV}
\newcommand*{\LM}{LIGO - Massachusetts Institute of Technology, Cambridge, MA 02139, USA}
\affiliation{\LM}
\newcommand*{\LU}{Louisiana State University, Baton Rouge, LA  70803, USA}
\affiliation{\LU}
\newcommand*{\LE}{Louisiana Tech University, Ruston, LA  71272, USA}
\affiliation{\LE}
\newcommand*{\LL}{Loyola University, New Orleans, LA 70118, USA}
\affiliation{\LL}
\newcommand*{\MT}{Montana State University, Bozeman, MT 59717, USA}
\affiliation{\MT}
\newcommand*{\MS}{Moscow State University, Moscow, 119992, Russia}
\affiliation{\MS}
\newcommand*{\ND}{NASA/Goddard Space Flight Center, Greenbelt, MD  20771, USA}
\affiliation{\ND}
\newcommand*{\NA}{National Astronomical Observatory of Japan, Tokyo  181-8588, Japan}
\affiliation{\NA}
\newcommand*{\NO}{Northwestern University, Evanston, IL  60208, USA}
\affiliation{\NO}
\newcommand*{\RI}{Rochester Institute of Technology, Rochester, NY  14623, USA}
\affiliation{\RI}
\newcommand*{\RA}{Rutherford Appleton Laboratory, HSIC, Chilton, Didcot, Oxon OX11 0QX United Kingdom}
\affiliation{\RA}
\newcommand*{\SJ}{San Jose State University, San Jose, CA 95192, USA}
\affiliation{\SJ}
\newcommand*{\SM}{Sonoma State University, Rohnert Park, CA 94928, USA}
\affiliation{\SM}
\newcommand*{\SE}{Southeastern Louisiana University, Hammond, LA  70402, USA}
\affiliation{\SE}
\newcommand*{\SO}{Southern University and A\&M College, Baton Rouge, LA  70813, USA}
\affiliation{\SO}
\newcommand*{\SA}{Stanford University, Stanford, CA  94305, USA}
\affiliation{\SA}
\newcommand*{\SR}{Syracuse University, Syracuse, NY  13244, USA}
\affiliation{\SR}
\newcommand*{\PU}{The Pennsylvania State University, University Park, PA  16802, USA}
\affiliation{\PU}
\newcommand*{\UM}{The University of Melbourne, Parkville VIC 3010, Australia}
\affiliation{\UM}
\newcommand*{\MI}{The University of Mississippi, University, MS 38677, USA}
\affiliation{\MI}
\newcommand*{\SF}{The University of Sheffield, Sheffield S10 2TN, United Kingdom}
\affiliation{\SF}
\newcommand*{\TA}{The University of Texas at Austin, Austin, TX 78712, USA}
\affiliation{\TA}
\newcommand*{\TC}{The University of Texas at Brownsville and Texas Southmost College, Brownsville, TX  78520, USA}
\affiliation{\TC}
\newcommand*{\TR}{Trinity University, San Antonio, TX  78212, USA}
\affiliation{\TR}
\newcommand*{\BB}{Universitat de les Illes Balears, E-07122 Palma de Mallorca, Spain}
\affiliation{\BB}
\newcommand*{\UA}{University of Adelaide, Adelaide, SA 5005, Australia}
\affiliation{\UA}
\newcommand*{\BR}{University of Birmingham, Birmingham, B15 2TT, United Kingdom}
\affiliation{\BR}
\newcommand*{\FA}{University of Florida, Gainesville, FL  32611, USA}
\affiliation{\FA}
\newcommand*{\GU}{University of Glasgow, Glasgow, G12 8QQ, United Kingdom}
\affiliation{\GU}
\newcommand*{\MD}{University of Maryland, College Park, MD 20742 USA}
\affiliation{\MD}
\newcommand*{\AM}{University of Massachusetts - Amherst, Amherst, MA 01003, USA}
\affiliation{\AM}
\newcommand*{\MU}{University of Michigan, Ann Arbor, MI  48109, USA}
\affiliation{\MU}
\newcommand*{\MN}{University of Minnesota, Minneapolis, MN 55455, USA}
\affiliation{\MN}
\newcommand*{\OU}{University of Oregon, Eugene, OR  97403, USA}
\affiliation{\OU}
\newcommand*{\RO}{University of Rochester, Rochester, NY  14627, USA}
\affiliation{\RO}
\newcommand*{\SL}{University of Salerno, 84084 Fisciano (Salerno), Italy}
\affiliation{\SL}
\newcommand*{\SN}{University of Sannio at Benevento, I-82100 Benevento, Italy}
\affiliation{\SN}
\newcommand*{\SH}{University of Southampton, Southampton, SO17 1BJ, United Kingdom}
\affiliation{\SH}
\newcommand*{\SC}{University of Strathclyde, Glasgow, G1 1XQ, United Kingdom}
\affiliation{\SC}
\newcommand*{\WA}{University of Western Australia, Crawley, WA 6009, Australia}
\affiliation{\WA}
\newcommand*{\UW}{University of Wisconsin-Milwaukee, Milwaukee, WI  53201, USA}
\affiliation{\UW}
\newcommand*{\WU}{Washington State University, Pullman, WA 99164, USA}
\affiliation{\WU}

\author{}    \affiliation{\GU}    
\author{B.~P.~Abbott}    \affiliation{\CT}    
\author{R.~Abbott}    \affiliation{\CT}    
\author{R.~Adhikari}    \affiliation{\CT}    
\author{P.~Ajith}    \affiliation{\AH}    
\author{B.~Allen}    \affiliation{\AH}  \affiliation{\UW}  
\author{G.~Allen}    \affiliation{\SA}    
\author{R.~S.~Amin}    \affiliation{\LU}    
\author{S.~B.~Anderson}    \affiliation{\CT}    
\author{W.~G.~Anderson}    \affiliation{\UW}    
% ...

 \collaboration{The LIGO Scientific Collaboration, http://www.ligo.org}
 \noaffiliation

\begin{abstract}
We present an all-sky search for gravitational waves in the frequency range 1 
to 6 kHz during the first calendar year of LIGO's fifth science run.  This is
 the first untriggered LIGO burst analysis to be conducted above 3 kHz. We 
discuss the unique properties of interferometric data in this regime.  
161.3 days of triple-coincident data were analyzed. 
No gravitational events above threshold were observed and a frequentist upper
limit of 5.4 $\,\text{year}^{-1}$ on the rate of strong gravitational wave bursts was
placed at a 90\% confidence level.
Implications for specific theoretical models of gravitational wave emission 
are also discussed.
\end{abstract}

\maketitle

\section{Introduction}

LIGO (Laser Interferometer Gravitational Wave Observatory) \citep{ligo} is composed
 of three laser interferometers at two sites in the United States of 
America.  The interferometers known as H1, with 4 km arms, and H2, with 2 km arms, are colocated within 
the same vacuum system at the Hanford site in Washington state.  An additional 
4-kilometer-long interferometer, L1, is located in Louisiana's Livingston Parish.  
The detectors have similar orientation, as far as is possible
 given the curvature of the Earth's surface and the constraints of the sites on
 which they were built, in order to be sensitive to the same gravitational wave
 polarizations.  The relatively large separation between the two sites (approximately 3000 km) helps distinguish an actual 
gravitational wave appearing in both detectors from local environmental 
disturbances, which should not have a corresponding signal at the other site. 
GEO\,600, a 600 m interferometer located near Hannover Germany, also operates
as part of the LIGO Scientific Collaboration.

The fifth science run (S5) of the LIGO interferometers was conducted between 
November 2005 and October 2007.  LIGO achieved its design sensitivity during 
this run, roughly a factor of 2 improvement in sensitivity over the previous 
S4 run \citep{s4burst}.  Additionally, S5 was by far the longest science run 
and had the best duty cycle, collecting a full year of livetime of data with 
all 3 LIGO detectors in science mode.  This is an order of magnitude greater 
triple-coincident livetime than all previous LIGO science runs combined.  The 
analysis discussed in this paper uses data from the first calendar year of S5, 
covering data from 4 November 2005 to 14 November 2006.

\begin{figure}
\begin{center}
\mbox{
\includegraphics*[width=0.5\textwidth]{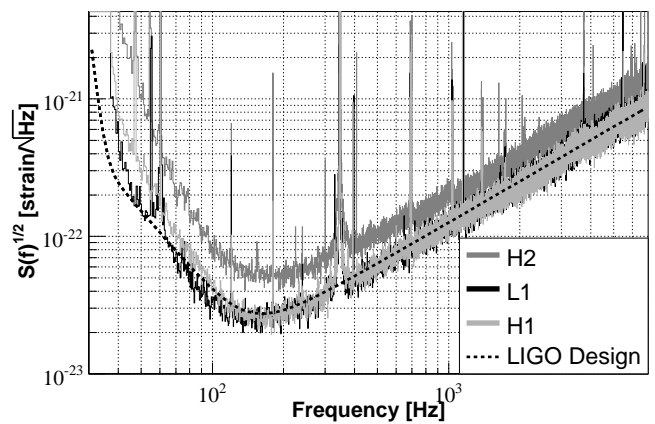}}
\caption{\label{fig:LIGOcurve} Characteristic LIGO sensitivity curves from June 2006. Shot noise dominates the spectrum at high frequencies.}
\end{center}
\end{figure}

Previous all-sky searches for bursts of gravitational waves with LSC 
instruments have been limited to frequencies below 3 kHz or less, in the range where 
the detectors are maximally sensitive \citep{s1burst,s2burst,s3burst,s4burst,s4burstgeo}.
  The sensitivity above 1 kHz is poorer than at lower frequencies because of 
the storage time limit of the interferometer arms, as demonstrated by 
the strain-equivalent noise spectral density curve 
for H1, H2 and L1 shown in Figure \ref{fig:LIGOcurve}.  Shot noise (random 
statistical fluctuations in the number of photons hitting the photodetector) is 
the dominant source of noise above $\sim$200 Hz.

Despite the higher noise floor, the interferometers are still sensitive enough 
to merit analysis in the few kilohertz regime and there are a number of models which 
lead to gravitational wave emission above 2 kHz.
As the sensitivity of gravitational wave interferometers continues to improve, 
it is important to explore the full range of data produced by them. 
LIGO samples data at 16384 Hz, in principle allowing analysis 
up to 8192 Hz, but the data are not calibrated up to the Nyquist frequency. 
Thus, this paper describes an all-sky high frequency search for gravitational 
burst signals using H1, H2 and L1 data in triple coincidence in the frequency range
1--6 kHz. This search complements the all-sky
burst search in the 64 Hz--2 kHz range, described in \cite{lowfreq}.

This paper is organized as follows: Section II describes the 
theoretical motivation for conducting this search.  Section III describes the 
analysis procedure.  Section IV discusses general properties of high frequency 
data and systematic uncertainties.  Section V discusses detection efficiencies
 based on simulated waveforms.  Results are presented in section VI, followed 
by discussion and summary in section VII.

\section{Transient Sources of Few-kHz Gravitational Waves}

A number of specific theoretical models predict transient gravitational wave 
emission in the few-kilohertz range.  One such potential source of emission is 
gravitational collapse, including core-collapse supernova and long-soft gamma-ray 
burst scenarios \citep{OttBurrows} 
which are predicted to emit gravitational waves in a range extending above 1 
kHz. In a somewhat higher frequency regime 
are neutron star collapse 
scenarios resulting in rotating black holes \citep{brwave1,brwave2}.  

Another  potential class of high frequency gravitational wave sources is 
nonaxisymmetric hypermassive neutron stars resulting from neutron star-neutron 
star mergers.  If the equation of state is sufficiently stiff, a 
hypermassive neutron star is formed as an intermediate step during the merger 
of two neutron stars before a final collapse to a black hole, 
whereas a softer equation of state leads to prompt formation of a black hole. 
Some models predict gravitational wave emission in the 2--4 kHz range from this
intermediate hypermassive neutron star, but in many cases higher frequency emission 
(6--7 kHz) from a promptly formed black hole \citep{HMNS,HMNS2}.  Observation of 
few-kilohertz gravitational wave emission from such systems would thus provide 
information about the equation of state of the system being studied. 

Other possible sources of few-kilohertz gravitational wave emission include neutron 
star normal modes (in particular the f-mode) \citep{fmodes} as well as neutron 
stars undergoing torque-free precession as a result of accreting matter from a 
binary companion \citep{nsm}. Low-mass black hole mergers \citep{lmbh}, soft gamma repeaters
\citep{SGR} or some scenarios for gravitational emission 
from cosmic string cusps \citep{cosmicstring} are additional possible sources. 
The majority of predicted high frequency gravitational wave signals tend to be 
of a few cycles duration in most scenarios since strong signals tend to lead to strong 
backreactions and hence significant damping. 

While there are specific waveform predictions from many of these models (some of which
 are studied in this analysis) these models still have substantial uncertainties and are only 
valid for systems with very specific sets of properties (e.g. mass and spin).
  Thus, as has been done previously for lower frequencies in each science run, 
we use search techniques that do not make use of specific waveforms. We require only short 
 ($\ll$1 s) duration and substantial signal power in the analysis band.

\section{Data Analysis}

\begin{figure}
\begin{center}
\mbox{
\includegraphics*[width=0.5\textwidth]{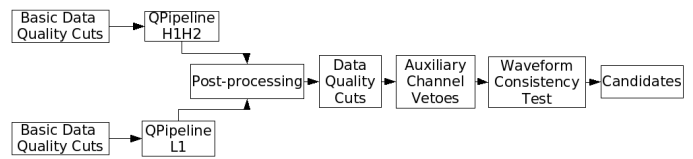}}
\caption{\label{fig:pipeline}A schematic of the analysis pipeline.  \emph{Triggers}, 
which are times when the power in one or more interferometer's readout
is in excess of the baseline noise, 
are generated using the QPipeline algorithm \citep{lowfreq,shourovthesis,
qtransform}.  Post-processing includes checking for a corresponding trigger at 
the other site and clustering remaining triggers into 1 second periods to avoid
 multiple triggers from the same source.  Data quality cuts remove triggers 
from times with known disturbances which can contaminate the data with 
spurious transients of mundane origin. The remaining triggers are subjected to 
auxiliary channel vetoes and finally a waveform consistency test is performed
 using CorrPower \citep{CorrPower}.}
\end{center}
\end{figure}

The process of identifying potential gravitational wave candidate events and 
separating them from noise fluctuations and instrumental glitches takes place 
in several steps. A schematic outline of the analysis pipeline is shown in 
Fig.~\ref{fig:pipeline}. Using whitened data, triggers with frequency above 1kHz
are identified separately at the two LIGO sites 
using the QPipeline algorithm \citep{lowfreq,shourovthesis,qtransform}, then 
combined with triggers of consistent time and frequency at the other site in 
the post-processing stage.  The data quality cuts and veto stages remove
 triggers correlated with instrumental and environmental disturbances that are known 
to be not of gravitational wave origin. 
Remaining triggers are then subjected to a final cut based on the consistency
 of the signal shape in the three interferometers \citep{CorrPower}. The 
analysis procedure is described in greater detail in the remainder of this 
section.  


These procedures were developed using time-shifted data produced by 
sliding the time stamps of Livingston triggers relative to Hanford triggers 
with  100 different time-shifts in increments of 5 seconds. Applying multiple 
time-shifts allows us to produce a set of independent time-shifted triggers 
with an effective livetime much larger than the actual livetime of the analysis.  Since 5 
seconds is much longer than the light travel time between the detectors, even after padding
for the finite time resolution of our search, no genuine gravitational wave signals will be coincident 
with themselves in the time-shifted data streams, allowing us to use this set of time-shifted triggers 
as background data. H1 and H2 data streams are not shifted relative to
 each other because their common environment is likely to produce temporally 
correlated non-stationary noise, meaning that time-shifts between H1 and H2
 would not accurately represent real background.  The analysis was tested on a 
single day of data (December 11th, 2005), then extended to the 
entire first calendar year of S5.  

GEO\,600, a 600 meter interferometer in Germany, was also collecting gravitational 
wave data during this timeframe. However, since the smaller GEO\,600 interferometer
 is substantially less sensitive than LIGO, including GEO\,600 would not 
have caused a substantial increase in overall sensitivity.  Also, incorporating
 an additional interferometer not co-aligned with the others would have added 
substantial complications to the analysis, especially since the  
cross-correlation test we perform with CorrPower \citep{CorrPower} is not 
designed to analyze data from detectors that are misaligned.  Thus, for this
analysis, we used GEO\,600 data as a follow-up only, to be examined in the case that
 any event candidates were identified using LIGO.  Virgo \citep{virgo}, a 3 km interferometer 
located in Cascina, Italy, was not operating during the period described in this paper.
Joint analysis of LIGO and Virgo data at high frequencies will be described in a future publication.

\subsection{The QPipeline Algorithm}

The QPipeline algorithm is run on calibrated strain data \citep{ht calibration}
 to identify triggers.  Each trigger is identified by a central time, duration, central frequency, bandwidth  and normalized energy.  Any trigger surviving to the end of the pipeline described in 
Figure \ref{fig:pipeline} would be considered as a gravitational wave candidate event. However, 
the vast majority of triggers generated by QPipeline are of mundane 
origin.  

Before searching for triggers, QPipeline whitens the data using zero-phase linear predictive
filtering \citep{shourovthesis,lpf2}.  In linear predictive filtering, a given sample in a 
data set is
assumed to be a linear combination of $M$ previous samples.  A modified zero-phase whitening
filter is constructed by zero-padding the initial filter, converting to the frequency domain
and correcting for dispersion in order to avoid introducing phase errors \citep{lowfreq}.

QPipeline is based on the Q transform, wherein the time series $s(t)$ is 
projected onto complex exponentials with bisquare windows, defined by central 
time $\tau$, central frequency $f_{0}$ and quality factor \emph{Q} (approximately the number of cycles present 
in the waveform).  
This can be represented by the formula


\begin{equation}
\begin{array}{rcl}
X(\tau,f_0,Q)  & = & \int_{-\infty}^{+\infty} \tilde{s}(f)\left(\frac{315}{128 \sqrt{5.5}}\frac{Q}{f_0}\right)^{1/2} \\ & & \left[1-(\frac{f Q}{f_{0} \sqrt{5.5}})^2\right]^2 e^{+i2\pi f \tau} df .
\end{array}
\end{equation}

Because it uses a set of generic complex exponentials as a template bank, 
QPipeline thus functions much like a matched filter search for waveforms which 
appear as sinusoidal Gaussians after the data stream is whitened \citep{shourovthesis} .  
This bank of templates is tiled logarithmically in $Q$ and frequency, but tiles at a given frequency
are spaced linearly in time.  The 
templates are spaced in such a way that we lose no more than 20\% of the trigger's normalized 
energy due to mismatches $\delta t$, $\delta f$ and $\delta Q$ \citep{lowfreq}.

The significance of a trigger is expressed in terms of its normalized energy $Z$, 
defined by taking the ratio of the squared projection magnitude to the mean squared projection
magnitude of other templates at the same $Q$ and frequency:

\begin{equation}
Z = |X|^2/\langle|X|^2\rangle \,.
\end{equation}
 

A gravitational wave signal would appear identical (in units of
calibrated strain) in the colocated, co-aligned H1 and H2 detectors
at the Hanford site.  Therefore, a new \emph{coherent} data stream is
formed from the noise-weighted sum of the two data streams.
 Mathematically, this can be expressed as:

\begin{equation}
\tilde{s}_{H+} = (\frac{1}{S_\text{H1}}+\frac{1}{S_\text{H2}})^{-1}(\frac{\tilde{s}_{\text{H1}}\left(f\right)}{S_\text{H1}}+\frac{\tilde{s}_{\text{H2}}(f)}{S_\text{H2}})
\end{equation}

\noindent where $S_\text{H1}$ and $S_\text{H2}$ are the power spectral densities of the two interferometers and
$\tilde{s}_{\text{H2}}\left(f\right)$ and $\tilde{s}_{\text{H1}}\left(f\right)$ are the frequency domain representation
of the strain data coming from H1 and H2.


The coherent analysis also defines a \emph{null} stream, $H_-$, which is just the 
normalized difference between the strain data of H1 and H2.  For lower frequency analyses, 
if the null $H_-$ stream value is too large the coherent $H_+$ stream is 
vetoed at the corresponding time~\citep{lowfreq}.  This is because a signal 
with consistent magnitude in both detectors should cancel out to zero, so a 
large null stream value indicates an inconsistent signal detected by the two 
interferometers.  However, we do not apply this null stream consistency veto 
in the high frequency search and simply take the result of the coherent stream 
as the final QPipeline result for the Hanford site, leaving this consistency test as
part of the follow-up procedure to vet any gravitational wave candidates.  
This is for two reasons: 
a.) at the time the analysis was designed it was feared that
substantially larger systematic uncertainties in calibration at higher 
frequencies mean that the criterion for what constitutes \emph{consistent} 
behavior between the two Hanford detectors would have to have been 
substantially relaxed, and b.) a smoother, less glitchy background population
 makes this consistency test only marginally useful (less than a 1\% reduction
 in the clustered coincident background trigger rate) above 1 kHz in any case.

For this analysis, we threshold at a normalized energy \begin{math}Z=16\end{math} for both sites.  
Along with CorrPower $\Gamma$ (defined in section III E) this is one of the 
variables used to tune the false alarm rate of the analysis.  In the case of 
Livingston, $Z$ is simply the normalized energy coming out of the Q transform,
 whereas in the case of Hanford, this is the normalized energy coming out of 
the coherent stream.

While lower frequency data are analyzed at 4096 Hz to save on computational 
costs, this search needs the full LIGO rate of 16384 Hz in order to
analyze higher frequencies. This higher sampling rate required computational 
tradeoffs relative to lower frequency analysis. Specifically, data were 
analyzed in blocks of 16 seconds rather than 64 seconds due to memory 
constraints. Additionally, the templates applied covered signals with {\it Q} 
from 2.8 to 22.6 rather than extending to higher {\it Q}s in order to reduce 
the required processing 
time.  This choice of {\it Q} range is consistent with theoretical predictions,
 since the  models under study in this frequency range generally predict 
signals of a few cycles.  More detailed information on QPipeline can be found 
in \cite{lowfreq} and \cite{shourovthesis}.

\subsection{Post-Processing of Triggers}

After triggers have been identified at both sites, the two lists are 
combined into one coincident trigger list.  In order to form a coincident 
trigger, there must be triggers at both sites which have time and frequency 
values consistent with each other.  Specifically, the peak 
times $\tau_{H}$ and $\tau_{L}$ at the Hanford and Livingston sites must satisfy the inequality

\begin{equation}
|\tau_H - \tau_L| < \mathrm{max}(\delta\tau_H , \delta\tau_L)/2 + 20~\mathrm{ms}
\end{equation}

\noindent where $\delta\tau_{H}$ and $\delta\tau_{L}$ are the durations of the two 
triggers.  The time of flight for a gravitational wave traveling directly 
between the detectors is approximately 10 ms, so a 20 ms coincidence window is 
somewhat padded to allow for misreconstructions in the central time of the 
waveform.  This is a more conservative window choice than that of the 
corresponding QPipeline S5 all-sky burst search at lower frequencies 
\citep{lowfreq}, but the difference in the coincidence window has minimal 
effect on the sensitivity of the analysis.

Similarly, the central frequencies $f_{0,H}$ and $f_{0,L}$ of the two triggers must 
satisfy the condition:

\begin{equation}
|f_{0,H} - f_{0,L}| < \mathrm{max}(\delta f_{0,H} , \delta f_{0,L})/2.
\end{equation} 

\noindent where $\delta f_{0,H}$ and $\delta f_{0,L}$ are the bandwidths of the triggers at the two 
sites.  This definition is identical to that used 
in the lower frequency analysis.

Once this coincident list has been obtained, the coincident triggers are clustered in 
periods of 1 second, taking only the trigger with the highest normalized energy, in order 
to eliminate multiple triggers from the same feature in the data stream.  
The remaining downselected triggers are referred to as \emph{clustered triggers}.

\subsection{Data Quality Cuts}

\emph{Data quality cuts} are designed to remove periods of data during which there is 
an unusually high rate of false triggers due to known causes.
  An effective data quality cut should remove a large number of spurious 
background triggers while resulting in a relatively small reduction in the 
livetime of the analysis.  These cuts are selected from a predetermined set of data quality 
\emph{flags}, which identify times in which environmental monitors suggest a disturbance 
that might influence the gravitational-wave readout.  The determination of which data quality 
flags to apply is made based on single detector properties and an exact procedure for 
application of these flags is put in place before generating coincidences which may be 
considered gravitational wave candidates.  The application of data quality flags therefore 
does not affect the statistical validity or ``blindness'' of the search.  We use the same 
\emph{category 1} and \emph{category 2} 
data quality cuts as the S5 low frequency burst searches \citep{lowfreq}.  
Category 1 cuts remove periods of time where there were major, obvious 
problems, such as a calibration line dropout or the presence of hardware 
injections, which make the data unusable.  Similarly, category 2 
cuts remove periods for which there is a clear external disturbance which 
distorts the data.  Category 2 cuts result in a loss of 1.4\% of the 
triple-coincident livetime.  While category 1 periods are removed before the 
start of the analysis, category 2 periods are removed at a later stage so as to
 avoid creating a large number of very short science segments which are 
impractical to process using QPipeline.

\emph{Category 3} data quality flags, 
which define periods where the data are analyzable but still somewhat suspect 
due to some known cause, were studied one at a time for their effectiveness 
relative to high frequency triggers. The category 3 flags used for 
this high frequency analysis are a subset of those adopted at low frequencies.
  Flags which removed QPipeline background triggers at a much higher rate 
than expected by random Poisson coincidence were selected for use.  
Specifically, the rate of clustered single site triggers must be at least 1.7 times higher 
for periods when a given data quality flag is on relative to periods when that 
flag is off.  As in the lower frequency analyses, category 3 data quality flags
 are used only for purposes of setting the upper limit, but triggers surviving 
to the end of the pipeline may still be examined as gravitational wave event 
candidates if they are within a category 3 data quality segment. The flags used
 in this high frequency analysis are summarized in Table \ref{table:dq3}.  
Applying the selected category 3 data quality flags ultimately removes 19.4\% 
of the surviving coincident time-shifted background triggers and results in a 1.7\% 
reduction in triple coincident livetime.

\begin{table*}
\caption{\label{table:dq3}Category 3 data quality cuts for high frequency analysis.}
\begin{ruledtabular}
\begin{tabular}{|l|r|r|r|}
          &             & Livetime & Ratio of Clustered Trigger Rate\\
Flag name & Description & Loss (s) & (Flag On:Flag Off)\\
\hline
H1:WIND\_OVER\_30MPH & Heavy wind at & 5531  & 1.93\\
 &ends of H1 arms & &\\
\hline
H1:DARM\_09\_11\_DHZ\_HIGHTHRESH &Up-conversion of seismic& 6574  & 1.76 \\
 &noise at 0.9 to 1.1 Hz& &\\
\hline
H1:SIDECOIL\_ETMX\_RMS\_6HZ &Saturation of side coil& 1360 & 2.11 \\
 &current in H1 X end mirror& &\\
\hline
H1:LIGHTDIP\_02\_PERCENT & Significant dip in stored& 34336 & 2.24 \\
&laser light power in H1& &\\
\hline
H2:LIGHTDIP\_04\_PERCENT & Significant dip in stored& 40562 & 2.04 \\
&laser light power in H2& &\\
\hline
L1:LIGHTDIP\_04\_PERCENT & Significant dip in stored& 115584 & 2.85 \\
&laser light power in L1& &\\
\hline
L1:BADRANGE\_GLITCHINESS & Abrupt drop in interferometer& 3185 & 1.95 \\
&sensitivity, quantified in terms of& &\\
&effective range for inspiral signals& &\\
\hline
L1:HURRICANE\_GLITCHINESS & Hurricane was active near Livingston& 42917 & 2.92\\
\hline
\end{tabular}
\end{ruledtabular}
\end{table*}

\subsection{Auxiliary Channel Vetoes}

The LIGO interferometers use a large set of auxiliary detectors to determine 
when potential event candidates are the result of environmental causes (such as
 seismic activity or electromagnetic interference) or problems with the 
interferometer itself rather than actual gravitational waves. Triggers from these
auxiliary detectors act as \emph{vetoes}, removing potential gravitational wave
candidate events that occur at the same time as the trigger in the auxiliary detector.
These vetoes are distinguished from the data quality cuts described in the previous section
because they are determined in a statistical way and remove triggers from a much 
shorter period of time (tens to hundreds of milliseconds around a particular veto trigger 
rather than blocks of seconds to thousands of seconds in the case of data quality cuts).
As with the data quality flags described above, all tuning of event-by-event vetoes is done
on a single instrument basis before coincident triggers are generated.
Vetoes are divided into categories using the same definitions as data quality flags.
The same list of category 2 vetoes used at low frequencies \cite{lowfreq} was applied to this search.
These vetoes require multiple magnetometer or seismic 
channels at a given site to be firing simultaneously.

This analysis also uses the same method of selecting which category 3 auxiliary channel vetoes 
to apply as was used for the lower frequency S5 all-sky searches, but used an independent set of 
high frequency QPipeline time-shifted background triggers to select these vetoes.  
A list of potential vetoes is assembled from the various auxiliary channels at different
thresholds and with different coincidence windows.  The effectiveness of each potential veto 
is measured by its efficiency-to-deadtime ratio, which is the percentage of background
triggers it removes from the analysis divided by the percentage of the total livetime it removes.
The vetoes which are actually applied are selected in a hierarchical fashion, first picking
 the most effective veto, then calculating the effectiveness of the remaining 
possible vetoes after this one has been applied.  The next most effective veto 
is then selected and the process repeated until all remaining veto 
candidates have either an efficiency-to-deadtime ratio less than 3 or a 
probability of their effect resulting from random Poisson coincidence greater 
than 10$^{-5}$.  The vetoes were selected using a set of background triggers 
obtained from 100 time-shifts of L1 with respect to H1H2, with offsets ranging 
from -186 to 186 seconds in increments of 3 seconds. Time-shifts which were also 
divisible by 5 and thus present in the set used to determine the final 
background of the analysis were omitted, making the veto training and test 
sets independent. Of 18831 triggers remaining in time-shifted background after
 category 3 data quality cuts, 2284 are removed by vetoes (12\% efficiency), 
while the vetoes cause a 2\% reduction in the overall livetime of the analysis.

\subsection{Cross-Correlation Test with CorrPower}

The remaining clustered triggers are next subjected to cross-correlation 
consistency tests using the program CorrPower \citep{CorrPower}. CorrPower has 
previously been used in S3 and S4 analyses \citep{s3burst,s4burst}.  Unlike 
QPipeline, which only looks for excess power on a site-by-site basis, CorrPower
 thresholds on normalized correlation between data streams in different 
detectors. CorrPower was selected for use in this analysis because it is 
relatively fast computationally and effective for roughly co-aligned 
interferometers such as LIGO.  For analyses including detectors with 
substantially different alignments relative to LIGO, such as Virgo or GEO\,600, one 
does not necessarily obtain consistent correlated signals between 
interferometers and more sophisticated fully coherent techniques such as 
Coherent WaveBurst \citep{cwb} or X-pipeline \citep{xpipeline} would be 
preferable.

Before applying the correlation test, data was filtered to the 1--6 kHz target frequency range
of the search.  Additionally, triggers 
were rejected entirely if their central frequency as determined by QPipeline 
was greater than 6 kHz.  Since this analysis extends CorrPower to higher 
frequency regimes compared to previous analyses, it was necessary to add 
\begin{math}Q=400\end{math} notch filters at frequencies of 3727.0, 3733.7, 5470.0 and 5479.2 
Hz, which correspond to ``butterfly'' and ``drumhead'' resonant frequencies of 
the interferometers' optical components.  The data are whitened. CorrPower then measures 
correlation using Pearson's linear correlation statistic:

\begin{equation}
r= \frac{\sum_{i=1}^N (x_i-\bar{x})(y_i-\bar{y})}{\sqrt{\sum_{i=1}^N (x_i-\bar{x})}\sqrt{\sum_{i=1}^N (y_i - \bar{y})^2}}
\end{equation}

\noindent where $x$ and $y$ are in this case the time series being compared for the two interferometers, 
$\bar{x}$ and $\bar{y}$ are the average values and N is the number of samples within the window 
used for the calculation.  This $r$-\emph{statistic} is calculated over windows of duration 
10, 25 and 50 ms.
  This variable is maximized over various time-shifts between the two interferometers.  The 
maximum time-shift between one of the Hanford detectors with the detector at Livingston is 11 ms, 
whereas the maximum time-shift between the two Hanford detectors is 1 ms.  The final output
 of CorrPower which we use as a data selection criterion is called $\Gamma$.  
$\Gamma$ is an average of the $r$-statistic values for each of the 3 detector
 combinations, using the integration length and relative time-shift between 
interferometers which results in the highest overall $r$-statistic value.  

\subsection{Tuning for the Final Cut}

CorrPower was run on the triggers resulting from the 100 background time-shifts.
This distribution was used to determine the value of the cut on the CorrPower 
$\Gamma$ output variable.  In order to obtain an estimated false alarm rate 
(FAR) of around one tenth of an event candidate in the analysis of 
time-shift-free foreground data, cuts were applied to remove the bulk of the 
time-shifted background distribution, only keeping 
triggers with $\Gamma$ values greater than 6.2 and a Qpipeline normalized energy greater than
$Z=16$ at both sites. This results in a final false alarm rate of 
$\sim 10^{-8}$ Hz.

\section{Properties of LIGO Data Above 1 kHz}

\subsection{High Frequency Trigger Distributions}

\begin{figure}
\begin{center}
\mbox{
\includegraphics*[width=0.5\textwidth]{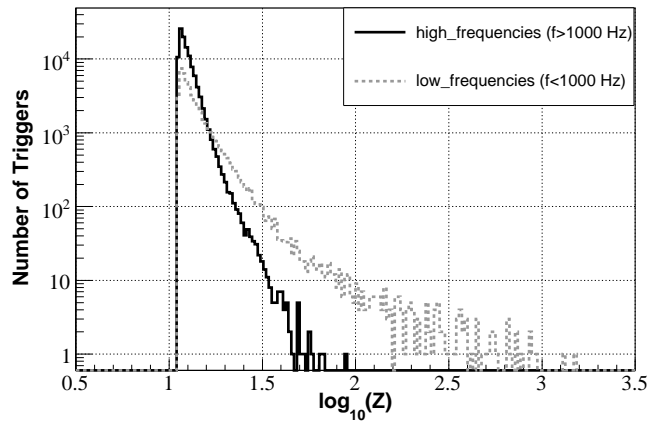}}
\caption{\label{fig:significance}Normalized energy Z of high and low frequency
 QPipeline triggers.  The low frequency distribution contains a substantially 
higher number of outliers.}
\end{center}
\end{figure}

Although the sensitivity of the detector is poorer at higher frequencies, the 
noise is more stationary in the shot-noise dominated regime. QPipeline 
normalized energy distributions from H1H2 for both high ($>$1 kHz) and low 
($<$1 kHz) frequency 
triggers are shown for a single day (December 11th, 2005) in Figure 
\ref{fig:significance}.  The distribution of single interferometer triggers at higher 
frequencies  falls off substantially more sharply than does the lower frequency
 distribution and contains far fewer statistical outliers.
The poorer statistics of the low frequency data set are due to glitches in the band
below 200 Hz.


\subsection{\label{sec:sysun}Systematic Uncertainties}
Due to variations in the response of the detectors as a function of frequency, systematic 
uncertainties are calculated separately for each of three detection bands: below 2 kHz, 2 to
4 kHz and 4 to 6 kHz.  The dominant source of systematic uncertainties is from
the  amplitude measurements in the frequency domain calibration. The individual 
amplitude uncertainties from each interferometer -- of order 10\% -- are combined into a single uncertainty 
by calculating a combined root-sum-square amplitude SNR and propagating the individual uncertainties in this 
equation assuming each error is independent.  In addition to this primary uncertainty, there is a small 
uncertainty 
(3.4\% or less depending on frequency band) introduced by converting from the frequency domain to the time domain strain series on which the analysis was actually run \citep{ht calibration}.

There is also phase uncertainty on the order of a few degrees in each interferometer and in each 
frequency band, arising both from the initial frequency domain calibration and the conversion to 
the time domain.  However, phase uncertainties are within acceptable tolerance.  In this analysis in 
particular, the omission 
of the null stream in QPipeline means the analysis is generally insensitive to phase shifts 
between the interferometers on the order of those observed.  Likewise, CorrPower
is mostly insensitive to phase shifts between interferometers because it automatically
maximizes over multiple time-shifts between the interferometers and will therefore
still find the maximum possible correlation. Some distortion in the shape of broadband
signals due to differing phase response at different frequencies is in principle possible.  However, 
this is not a significant concern since the phase uncertainties at all frequencies correspond to 
phase shifts on the order of less half a sample duration.  We therefore do not make any adjustment
to the overall systematic uncertainties due to phase error.

The antenna pattern for LIGO is normally calculated using the long wavelength 
approximation, which assumes the period of oscillation of a gravitational wave 
is large with respect to the transit time of a photon down the length of the 
interferometer arm and back.  This assumption is less accurate as the frequency  increases.  However, comparing results using the approximate 
long wavelength antenna pattern and frequency-dependent exact 
antenna pattern \citep{highfreqantpat} even towards the extreme high end of our
 frequency range (at 6 kHz) results in sensitivity calculations (see next 
section) differing by only $\sim$1\%.  Thus, the approximation of a constant antenna
 pattern has a negligible effect on the analysis.  Finally, we include a statistical uncertainty of around 2.7\% (with some variation from waveform to waveform
due to different numbers of injected waveforms).  

In each frequency band the frequency domain amplitude uncertainties 
are added in quadrature with the other smaller uncertainties to obtain the total uncertainty.  
The total 1 $\sigma$ uncertainties are then scaled by a factor of 1.28 to obtain the factor by which our $h_{\rm rss}$
limits are rescaled in order to obtain values consistent with 90\% confidence level upper limits.  These net uncertainty
values are 11.1\% in the less than 2 kHz band, 12.8\% in the 2-4 kHz band and 17.2\% in the 4-6 kHz band. 
Waveforms with significant signal content in multiple bands are considered to be in the band with the larger uncertainty.

\begin{figure}
\begin{center}
\mbox{
\includegraphics*[width=0.5\textwidth]{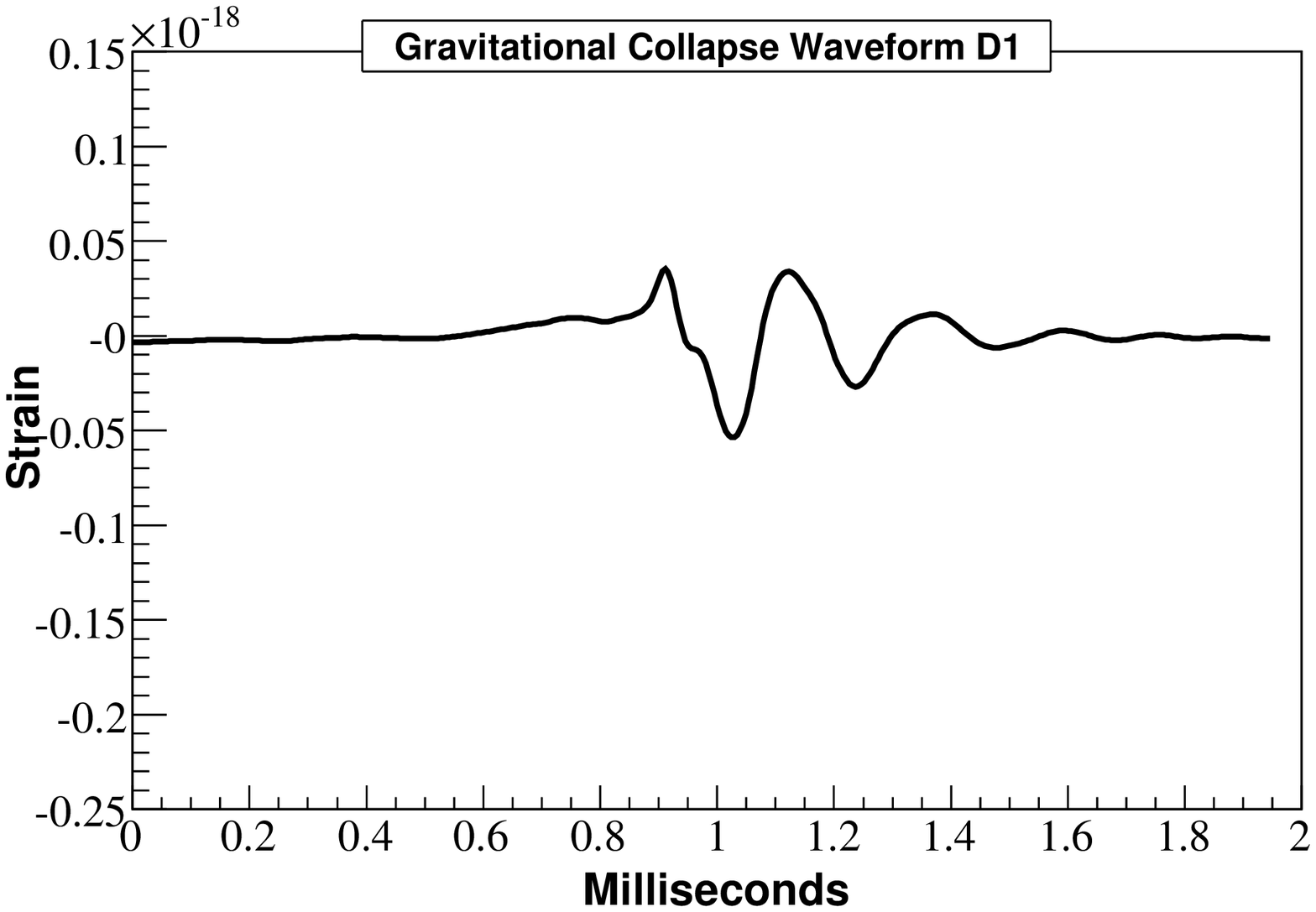}}
\mbox{
\includegraphics*[width=0.5\textwidth]{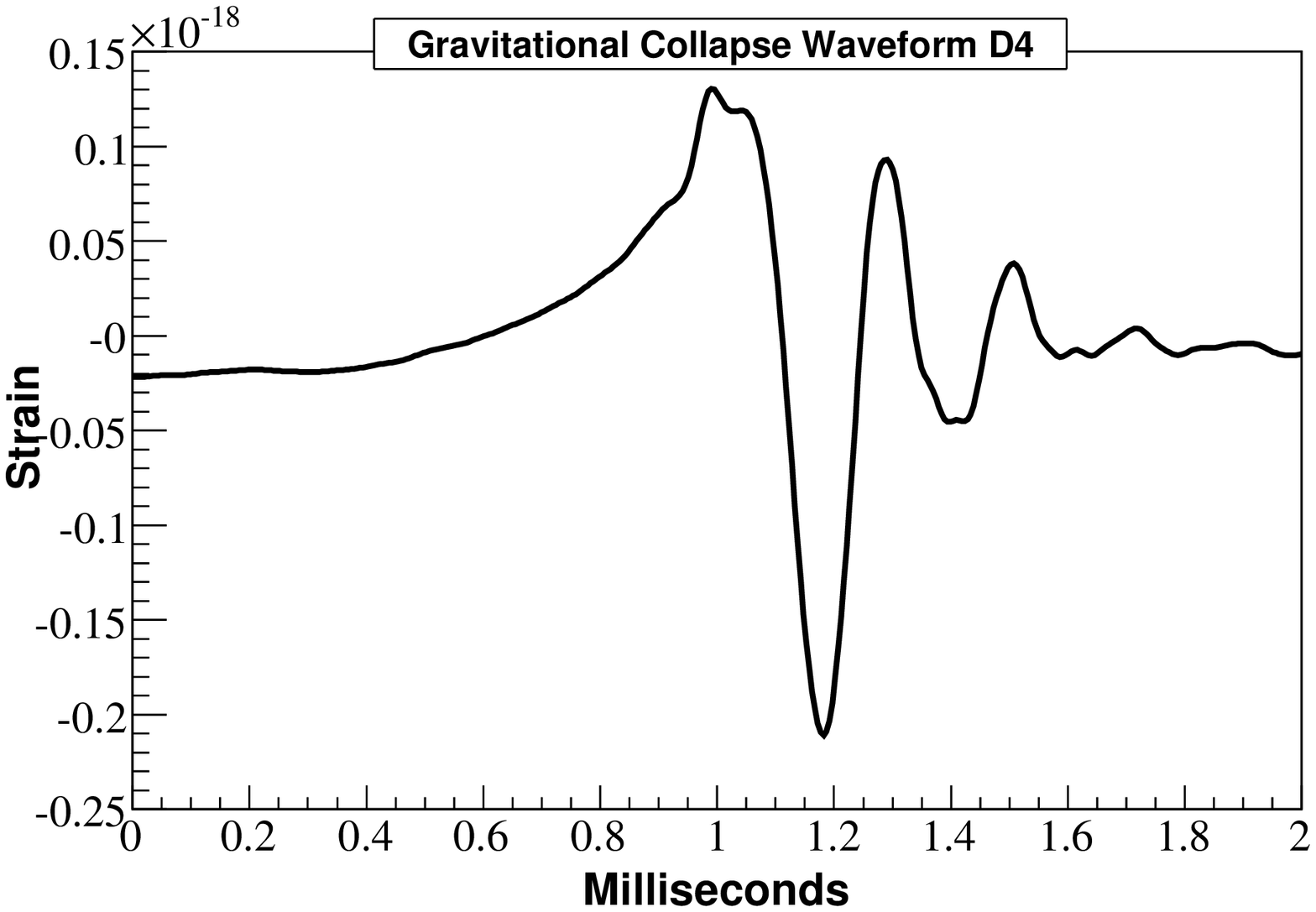}}
\caption{\label{fig:waveforms}Two example high frequency waveforms resulting 
from gravitational collapse of rotating neutron star models \citep{brwave2}.  
D1 results from a nearly 
spherical 1.26 solar mass star while D4 results from the collapse of a 
maximally deformed 1.86 solar mass star into a black hole.  The figures show 
the plus polarization for each waveform (the cross polarization is at least an 
order of magnitude weaker in both cases) at a distance of 1 kpc, assuming optimal
sky location and orientation.  At this distance, the $h_{\rm rss}$ magnitudes of the two
waveforms are $5.7 \times 10^{-22} \rm{Hz}^{-1/2}$ for D1 and 
$2.5 \times 10^{-21} \rm{Hz}^{-1/2}$ for D4.  They differ
from the figures presented in \citep{brwave2} in that the non-physical content
at the beginning of the simulations has been removed.}
\end{center}
\end{figure}

\section{Detection Efficiency}

Efficiency curves have been produced for three types of signal.  The cuts were developed on a set of 15 linearly polarized Gaussian-enveloped sine waves (sine-Gaussians) of the form

\begin{equation}
h\left(t_0 + t\right) = h_0 \sin\left(2\pi f_0t\right) \exp\left(-\left(2\pi f_0t\right)^2 /2Q^2\right)
\end{equation}

\noindent where $f_0$ and $t_0$ are the central frequency and time of the waveform and $Q$ is the
 quality factor defined previously. Additionally, we tested a set of three linearly polarized 
Gaussian waveforms as well as two waveforms taken from simulations by Baiotti {\it et al}. \cite{brwave2},
 which models gravitational wave emission from neutron star gravitational collapse and the ringdown of the
subsequently formed black hole using polytropes deformed by rotation.
The two scenarios studied here are designated D1, a nearly spherical 1.26 solar mass star, and D4, 
a 1.86 solar mass star that is maximally deformed at the time of its collapse into a black
 hole. These two waveforms are shown in Figure \ref{fig:waveforms}.  These two specific waveforms represent
 the extremes of the parameter space in mass and spin considered by Baiotti {\it et al}.



The BurstMDC and GravEn packages \citep{GravEn} were used to create simulated gravitational-wave
``injections'' which were superimposed on real data in a semi-random way
 at intervals of approximately 100 seconds.  This placed all injections far 
enough apart that whitening and noise estimation using data surrounding one injection is never
 affected by a neighboring injection.  Each waveform was 
simulated between 1000 and 1200 times for each of the 18 different amplitudes.  
 The intrinsic amplitude of a gravitational wave at the Earth, without folding in antenna
response factors, is defined in terms of its root-sum-squared strain amplitude:

\begin{equation}
\label{eqn:hrss}
h_\text{rss} \equiv \sqrt{\int{\left(|h_+\left(t\right)|^2 + |h_{\times}\left(t\right)|^2\right)}\rm{dt}}
\end{equation}

\noindent where $h_+\left(t\right)$ and $h_{\times}\left(t\right)$ are the plus and cross-polarization strain functions of the wave.  Since $h$ is a dimensionless quantity,  $h_{\rm rss}$ is given in
 units of Hz$^{-1/2}$.  

\begin{figure}
\begin{center}
\mbox{
\includegraphics*[width=0.5\textwidth]{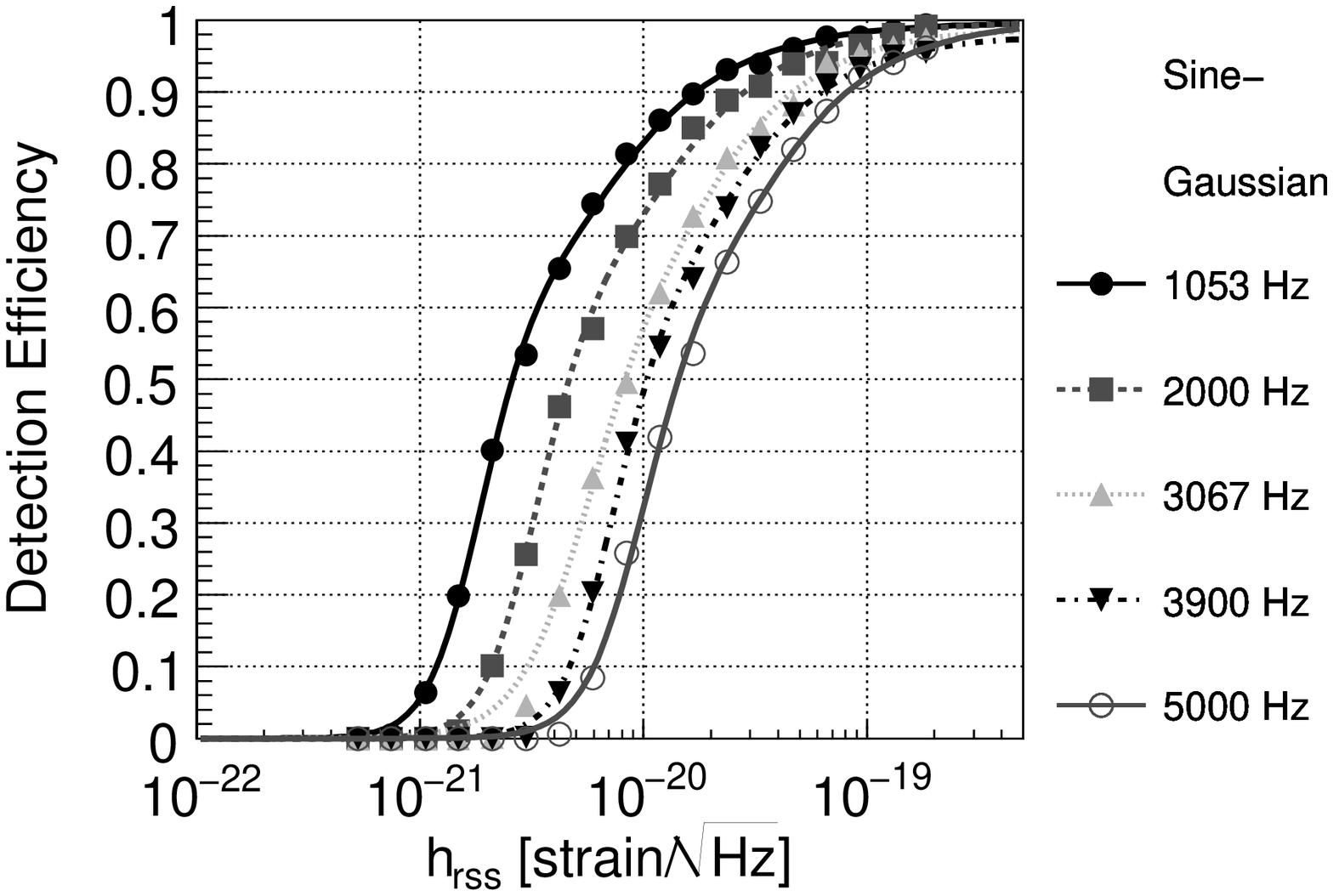}}
\mbox{
\includegraphics*[width=0.5\textwidth]{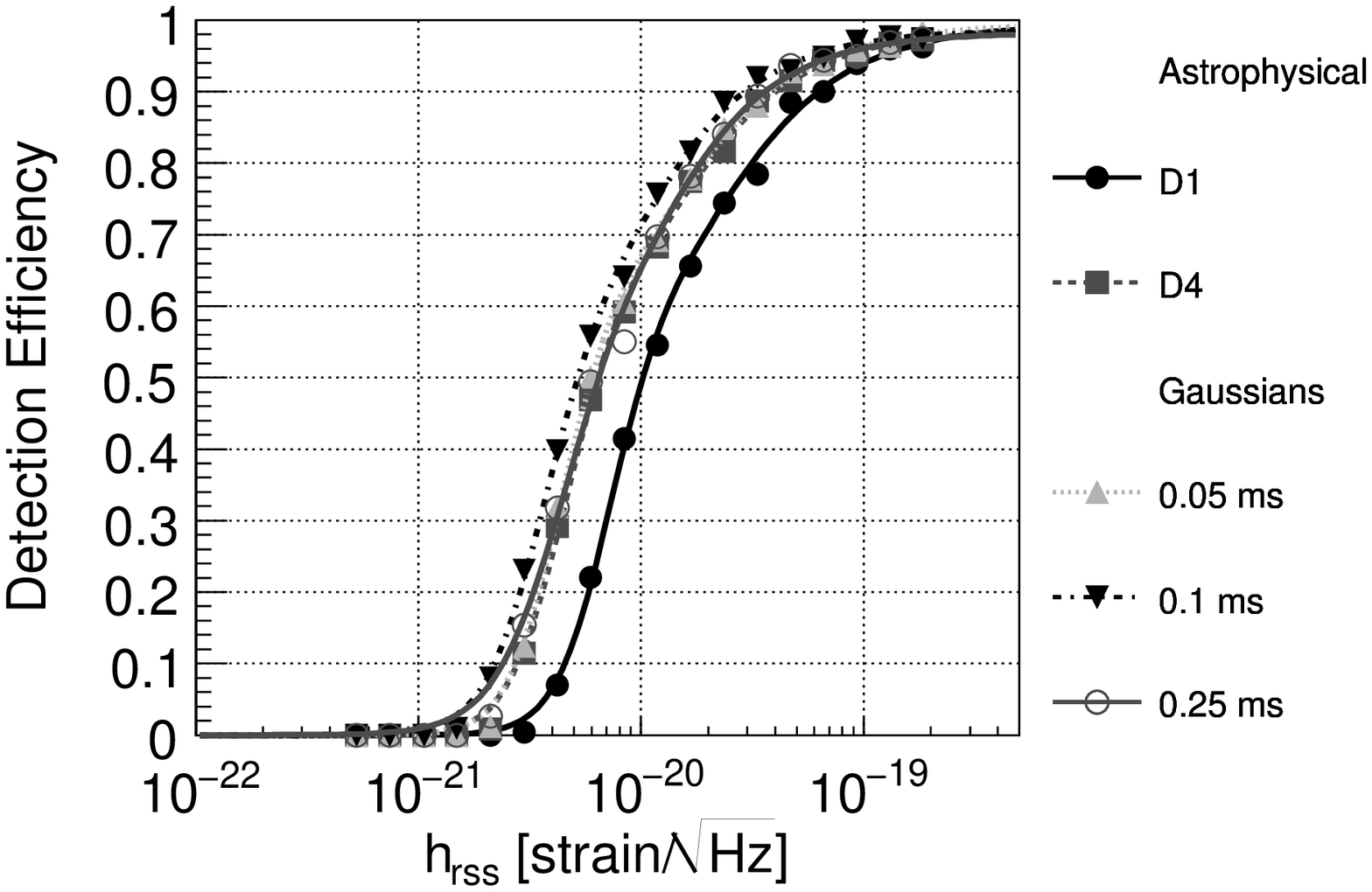}}
\caption{\label{fig:sgefficiency} Software injection efficiency curves for the 
set of sine-Gaussians of various frequencies (top) and Gaussians plus 
astrophysical waveforms (bottom).  There is a consistent reduction in 
efficiency as a function of frequency following the noise distribution.}
\end{center}
\end{figure}

\begin{figure}
\begin{center}
\mbox{
\includegraphics*[width=0.5\textwidth]{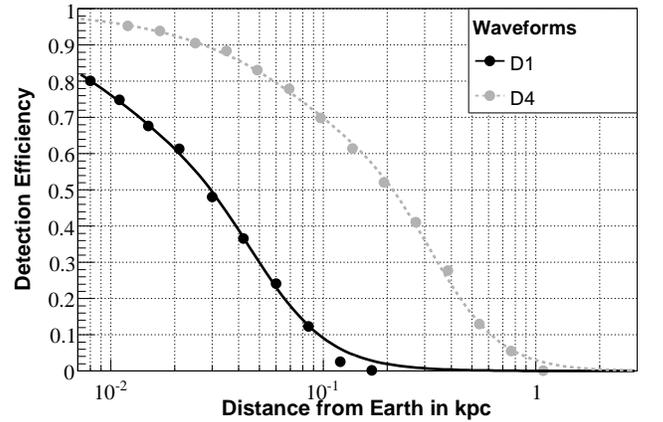}}
\caption{\label{fig:astrorange} Efficiency as a function of distance from Earth
 for supernova collapse waveforms D1 and D4 \citep{brwave2}, assuming random sky location, polarization and inclination angle $\iota$. A $\sin^2(\iota)$ dependence on the inclination angle was assumed.}
\end{center}
\end{figure}

The injections were distributed isotropically over the sky. Thus, even a few nominally 
very strong software injections are 
missed by the pipeline because they are oriented in a very sub-optimal way 
relative to at least one interferometer.
Since they are simulating an actual
astrophysical system, the D1 and D4 waveforms also include a randomized source
inclination
in addition to random sky location and polarization. A $\sin^2(\iota)$ 
dependence on the inclination angle was assumed. Figure 
\ref{fig:sgefficiency} shows 
efficiency curves for some of these waveforms as a function of signal 
amplitude. The $h_{\rm rss}$ values for which 50\% and 90\% of sine-Gaussian
injections are 
detected are summarized in Table \ref{table:hrss}.  Figure \ref{fig:astrorange}
 shows the detection efficiency for the simulated D1 and D4 Baiotti {\it et al}.
 models as a function of distance from Earth, indicating
that a neutron star collapse would have to happen nearby (within a kiloparsec) to be
detectable at our current sensitivity. 

Hardware injections, wherein actuators were used to physically simulate a 
gravitational wave in the interferometers by moving the optical components, 
were performed throughout S5. Although the numbers and variety of amplitudes 
were not sufficient to produce hardware injection efficiency curves, 
sine-Gaussian hardware injections at 1304, 2000 and 3067 Hz were reliably 
recovered using the high frequency search pipeline at amplitudes large enough
 that their detection is expected based on sensitivities determined by software injection efficiencies.  
Table \ref{table:hardwareinjections} shows 
the central frequency, amplitude and fraction of hardware injections detected.
For hardware injections, amplitude is given in terms of $h_{\rm rss,det}$, the root
sum square of the strain in the detector.  This is defined analogously to equation
\ref{eqn:hrss}, with $h_{+\rm{,det}}$ and $h_{\times\rm{,det}}$ in place of $h_{+}$
and $h_{\times}$.

\begin{table}
\caption{\label{table:hrss}$h^{50\%}_{\rm rss}$ and $h^{90\%}_{\rm rss}$ values
 (the root sum square strain at which 50\% or 90\% of injections are detected) 
for $Q=9$ sine-Gaussians.  Values in this table are adjusted for systematic uncertainties
as described in section IV B.}
\begin{ruledtabular}
\begin{tabular}{|l|r|r|}
Central Frequency & $h^{50\%}_{\rm rss}$ (Hz$^{-1/2}$) & $h^{90\%}_{\rm rss}$ 
(Hz$^{-1/2}$)\\
1053 & 2.87$\times$10$^{-21}$&1.97$\times$10$^{-20}$\\
1172 & 3.15$\times$10$^{-21}$&2.04$\times$10$^{-20}$\\
1304 & 3.31$\times$10$^{-21}$&2.06$\times$10$^{-20}$\\
1451 & 3.73$\times$10$^{-21}$&2.33$\times$10$^{-20}$\\
1615 & 3.99$\times$10$^{-21}$&2.67$\times$10$^{-20}$\\
1797 & 4.91$\times$10$^{-21}$&3.10$\times$10$^{-20}$\\
2000 & 5.22$\times$10$^{-21}$&3.30$\times$10$^{-20}$\\
2226 & 6.08$\times$10$^{-21}$&3.74$\times$10$^{-20}$\\
2477 & 6.63$\times$10$^{-21}$&4.47$\times$10$^{-20}$\\
2756 & 7.59$\times$10$^{-21}$&5.14$\times$10$^{-20}$\\
3067 & 9.20$\times$10$^{-21}$&5.62$\times$10$^{-20}$\\
3799 & 1.17$\times$10$^{-20}$&8.06$\times$10$^{-20}$\\
3900 & 1.19$\times$10$^{-20}$&7.87$\times$10$^{-20}$\\
5000 & 1.67$\times$10$^{-20}$&9.47$\times$10$^{-20}$\\
\end{tabular}
\end{ruledtabular}
\end{table}


\begin{table}
\caption{\label{table:hardwareinjections}S5 $Q=9$ sine-Gaussian hardware 
injections above 1 kHz. Note that $h_{\rm rss,det}$ and $h_{\rm rss}$ are different quantities because
$h_{\rm rss}$ does not include a sky location dependent antenna response factor, which will reduce the
detector response by an additional factor of 0.38 on average.  
Care should therefore be taken when comparing to Table I.}

\begin{ruledtabular}
\begin{tabular}{|l|r|r|}
Central Frequency (Hz) & $h_{\rm rss,det}$ (Hz$^{-1/2}$) & fraction 
recovered\\
1304 & 5.00$\times$10$^{-22}$& 0/2\\
1304 & 1.28$\times$10$^{-20}$& 16/16\\
1304 & 2.56$\times$10$^{-20}$& 16/16\\
2000 & 6.00$\times$10$^{-22}$& 0/102\\
2000 & 1.00$\times$10$^{-21}$& 0/4\\
2000 & 1.20$\times$10$^{-21}$& 14/127\\
2000 & 2.40$\times$10$^{-21}$& 125/125\\
2000 & 4.80$\times$10$^{-21}$& 117/117\\
2000 & 9.60$\times$10$^{-21}$& 21/21\\
2000 & 1.92$\times$10$^{-20}$& 16/16\\
2000 & 3.84$\times$10$^{-20}$& 16/16\\
3067 & 7.21$\times$10$^{-21}$& 13/13\\
3067 & 1.44$\times$10$^{-20}$& 13/13\\
3067 & 2.88$\times$10$^{-20}$& 3/3\\
3067 & 5.76$\times$10$^{-20}$& 3/3\\
\end{tabular}
\end{ruledtabular}
\end{table}

Good timing and frequency reconstruction help improve detection 
efficiency.  Using $Q=9$ sine-Gaussian waveforms, the timing resolution has been 
demonstrated to be within one cycle of the waveform and frequency resolution is
 better than 10\%, limited by the coarseness in frequency space of the templates used in QPipeline. 


\section{Results}

Having tuned the analysis on background from 100 time-shifts and tested it on a single day of data, we then performed the analysis on the actual 
coincident (or ``foreground'') data. No 
event candidates above our threshold were observed.  

\begin{figure}
\begin{center}
\mbox{
\includegraphics*[width=0.5\textwidth]{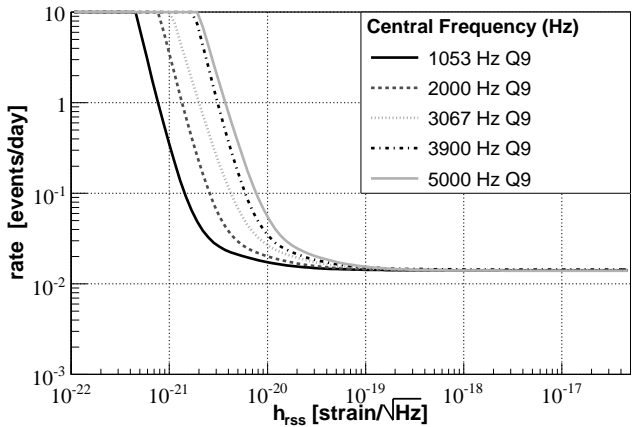}}
\caption{\label{fig:upperlimits} Upper limit curves for a number of our tested
 waveforms.  The rate at Earth of gravitational waves of each given type is excluded at
a 90\% confidence level.  The curves have been adjusted to account for systematic uncertainties 
as described in section IV B.}
\end{center}
\end{figure}

As in previous burst analyses (e.g. \cite{s4burst}), we set single-sided frequentist upper limits on the rate of gravitational wave emission.  
The upper limits in the frequency range 1--6 kHz are shown in Figure \ref{fig:upperlimits} for a sub-sample of our tested waveforms.
161.3 days of triple-coincident livetime were analyzed (see \cite{timelist} for
 a complete list of analyzed times). After performing predetermined category 
3 data quality cuts and vetoes, 155.5 days of triple-coincident data were used to set upper limits on gravitational wave emission. 

For gravitational waves with amplitudes 
such that detection efficiency approaches 100\%, the upper limit asymptotically
 approaches a value of 0.015 events per day (5.4 events per year), as determined 
primarily by the livetime of the analysis.  While other untriggered searches for 
gravitational waves with comparable or greater livetime (\emph{e.g.} the corresponding LIGO
lower frequency analysis \citep{lowfreq} and searches by IGEC \citep{IGEC}) have been conducted in overlapping frequency bands, this analysis
represents the first limit placed on gravitational wave emission over much of the frequency band.

\begin{table}
\caption{\label{table:counts}Number of triggers surviving various 
stages of the analysis: initial coincident triggers, triggers remaining after 
the removal of segments removed due to data quality criteria, triggers remaining
 after vetoes based on auxiliary channels have been applied and triggers 
ultimately surviving after the CorrPower linear correlation cut ($\Gamma$).  
Shown are results for 100 time-shifts, the same result normalized to the actual 
livetime, and the foreground results from the analysis performed without time-shifting the data. The background normalization reflects the fact that the livetime is different for different time-shifts.} 

\begin{ruledtabular}
\begin{tabular}{|l|r|r|r|}
 & Background & Normalized & Unshifted\\
 & Count & Background & Count\\
\hline
Coincident triggers &23361&242.9&265 \\
\hline
After data quality cuts &18831&195.8&223 \\
\hline
After auxiliary channel vetoes &16547&172.0&193 \\
\hline
After $\Gamma>$6.2 threshold &11 &0.115 &0 \\
\end{tabular}
\end{ruledtabular}
\end{table}

\begin{figure}
\begin{center}
\mbox{
\includegraphics*[width=0.5\textwidth]{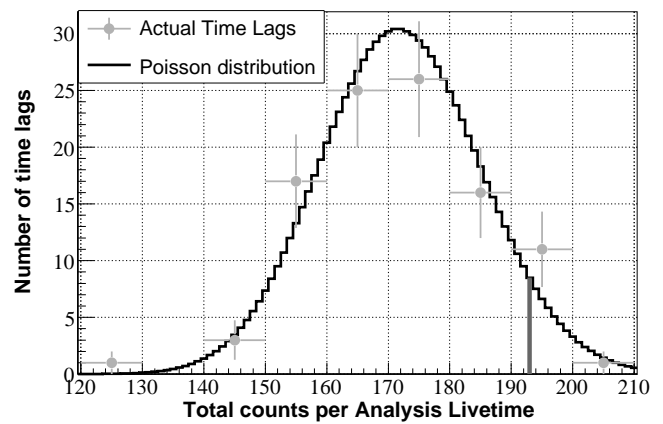}}
\caption{\label{fig:countshist} Histogram showing number of time-shifts vs. 
counts normalized to analysis livetime.  Superimposed is the expected 
distribution based on Poissonian statistics, which is consistent with the 
observed distribution.  The black line at 193 counts indicates the actual 
number of foreground triggers observed.}
\end{center}
\end{figure}

 The number of triggers surviving through each stage of the analysis are shown in 
Table \ref{table:counts}.  While there are no event candidates above our threshold in this analysis, the 
rates before the final CorrPower cut are slightly higher than expected.  However, assuming 
Poissonian statistics, this is not a statistically significant excess since there is a 6.2\% chance of 
getting at least the observed 193 foreground triggers after all data quality cuts and vetoes have been 
applied. Figure \ref{fig:countshist} demonstrates that the rate of 
triggers per time-shift can in fact be treated as a Poisson distribution.  



\begin{figure}
\begin{center}
\mbox{
\includegraphics*[width=0.5\textwidth]{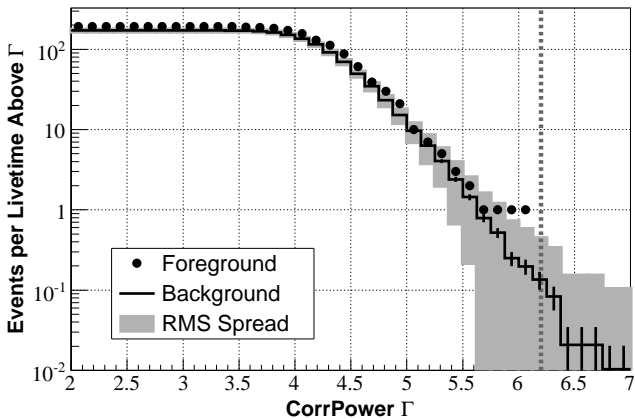}}
\caption{\label{fig:corrpower} CorrPower $\Gamma$ distribution for background 
(normalized to the livetime of the analysis) and foreground distributions 
before the final CorrPower cut.  The gray region is the RMS spread of counts in the 
background time-shifts while the error bars are the error in the mean counts per time-shift.
The dotted line shows the cut at $\Gamma$=6.2.}
\end{center}
\end{figure}

\begin{figure}
\begin{center}
\mbox{
\includegraphics*[width=0.5\textwidth]{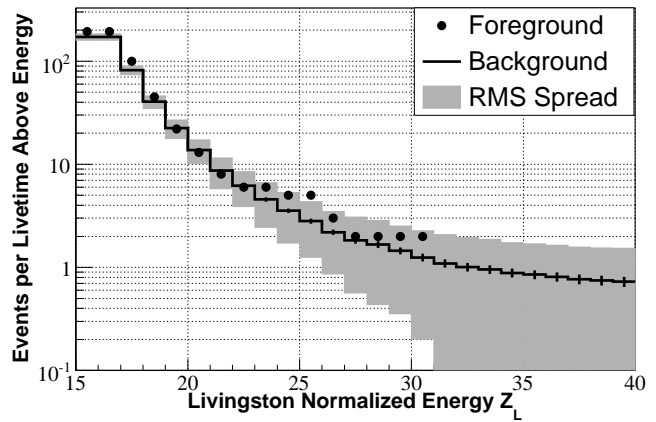}}
\mbox{
\includegraphics*[width=0.5\textwidth]{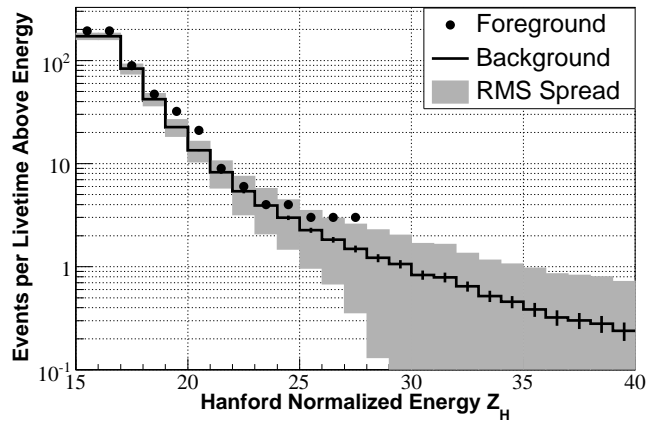}}
\caption{\label{fig:qsig} QPipeline significance distribution for background 
(normalized to the livetime of the analysis) and foreground distributions at 
the Livingston (top) and Hanford (bottom) sites before the final CorrPower cut.  The gray region is the RMS spread of counts in the background time-shifts while the error bars are the error in the mean counts per time-shift.}
\end{center}
\end{figure}

The foreground to background consistency of the CorrPower $\Gamma$ distribution
 (Fig. \ref{fig:corrpower}) and QPipeline normalized energies from the Livingston 
and Hanford sites (Fig. \ref{fig:qsig}) were also studied. These plots are 
produced after all data quality cuts and vetoes were applied, but 
before the final CorrPower $\Gamma$ cut.  The distributions are plotted 
cumulatively, i.e. each bin shows foreground and time-shifted background counts 
greater than or equal to the marked value. Other than the upward fluctuation in total 
counts already discussed, the distributions themselves are essentially 
consistent with expectation.

Since they appeared to stand out slightly from the expected background 
distribution (although not at a statistically significant level), the loudest 3
 triggers in Hanford QPipeline normalized energy, loudest 2 triggers in 
Livingston normalized energy and the trigger with highest CorrPower $\Gamma$ value were 
studied on an individual basis using Qscan 
\citep{shourovthesis}.  All of the triggers appear consistent with the background population.
In most cases the triggers arise from the correlation of a
fairly loud trigger with what appears to be one of a population of glitches of smaller magnitude 
in the other interferometers.  While only triggers passing category 3 data quality cuts were used 
to set the upper limit, the two events with the highest $\Gamma$ in the ``full'' data set after category 2 
cuts were also present after category 3. Since no triggers in the full data set were in apparent excess
of the stated upper limits, further follow-ups were not necessary.


In addition to the previously described search requiring data from all 3 LIGO
 interferometers, we also performed a check for interesting events during 
times in which H1 and H2 science quality data were available, but L1 data was 
not.  The two-detector search is less sensitive than the three-detector 
search and background estimation is less reliable, so we do not use this data when 
setting upper limits. 
However, in the first calendar year of S5, there are 77.2 days of livetime with 
only H1 and H2 data available (roughly half the livetime with simultaneous data
 from all three interferometers), so it is worth checking
 this data for potential gravitational wave candidates.  
This check used procedures similar to the analysis previously described, including 
identical data quality and veto procedures.  

\begin{figure}
\begin{center}
\mbox{
\includegraphics*[width=0.5\textwidth]{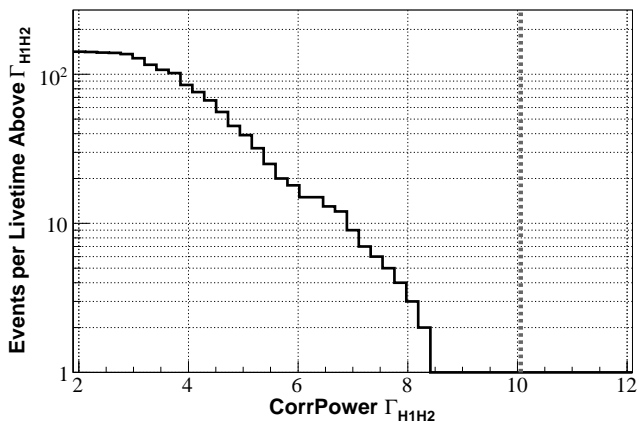}}
\caption{\label{fig:H1H2gamma} CorrPower Gamma zero-lag distribution for the H1H2 analysis. The dotted line shows the cut at $\Gamma$=10.1.}

\end{center}
\end{figure}

Due to the presence of correlated transients in H1H2 data, performing time-shifts of one
detector relative to the other is not a reliable means of obtaining an accurate background.
Instead, we use the unshifted H1H2 coincident triggers from the
H1H2L1 analysis as our estimate of the background since we have already determined that there are no
gravitational wave candidates in this data set. However, the H1H2L1 data
set is only about twice the livetime of the H1H2-only data set, so we are
required to extrapolate the false alarm probability distribution to
obtained the desired false alarm rate. To compensate for the uncertainties
in our estimate of the false alarm probability introduced by the reduced
data set and the extrapolation, we target a more conservative false alarm
probability of $\sim$0.01 triggers for the H1H2-only analysis.
This lower false alarm probability and the lack of L1 coincidence as a veto
requires stricter cuts, specifically coherent energy $Z>100$ from QPipeline and $\Gamma>10.1$ from CorrPower.  
As in the three-detector search, there were no events above threshold (see Fig. \ref{fig:H1H2gamma})
upon examination of the zero-lag foreground data, 
thus no potential gravitational wave candidates were identified in the two-detector search.

\section{Summary and Future Directions}

We have searched the few-kilohertz frequency regime for gravitational 
wave signals using the first calendar year of LIGO's fifth science run.  No 
gravitational wave events were identified, and we have placed upper limits on 
the emission of gravitational waves in this frequency regime. 

The second calendar year of S5 remains to be analyzed in this frequency range. 
Several months of this run overlap with the first science run of the Virgo 
\citep{virgo} detector, which began on May 18th, 2007.  During this period of 
overlap, data from Virgo as well as the LIGO interferometers will be 
incorporated into high frequency analysis.  Since Virgo is not co-aligned with the
LIGO detectors, this will require fully coherent analysis tools rather than CorrPower.
Above 1 kHz Virgo and LIGO have comparable sensitivities, making their combination 
especially advantageous in the few-kilohertz regime.  

The next LIGO science run will be done with Enhanced LIGO \citep{enhanced}, an improved version 
of the detectors.  Most relevant to high frequency analysis, the dominant 
background of shot noise will be reduced by increasing the power of the laser 
from 10 W to $\sim$35 W, substantially improving the sensitivity of the 
detectors.  Virgo+, a similarly enhanced version of Virgo, will operate 
simultaneously.  After this, further improvements will lead to the AdvancedLIGO
 \citep{advancedligo} and AdvancedVirgo \citep{advancedvirgo} detectors coming 
online around 2014.  Extending the analysis of gravitational wave data into the
 few-kilohertz regime will continue to be of scientific interest as these detectors 
become more and more sensitive.

\begin{acknowledgments}\label{sec:acknowledgements}

The authors gratefully acknowledge the support of the United States National
Science Foundation for the construction and operation of the LIGO Laboratory and
the Science and Technology Facilities Council of the United Kingdom, the
Max-Planck-Society, and the State of Niedersachsen/Germany for support of the
construction and operation of the GEO\,600 detector. The authors also gratefully
acknowledge the support of the research by these agencies and by the Australian
Research Council, the Council of Scientific and Industrial Research of India,
the Istituto Nazionale di Fisica Nucleare of Italy, the Spanish Ministerio de
Educaci\'on y Ciencia, the Conselleria d'Economia Hisenda i Innovaci\'o of the
Govern de les Illes Balears, the Royal Society, the Scottish Funding Council, 
the Scottish
Universities Physics Alliance, The National Aeronautics and Space
Administration, the Carnegie Trust, the Leverhulme Trust, the David and Lucile
Packard Foundation, the Research Corporation, and the Alfred P. Sloan
Foundation.  The authors thank Luca Baiotti and Luciano Rezzolla for providing 
simulation data and valuable discussion concerning the testing of astrophysical 
waveform models.

This document has been assigned LIGO Laboratory document number LIGO-P080080.

\end{acknowledgments}


\begin{thebibliography}{}


\bibitem[1]{ligo}
D.~Sigg for the LSC.  Class. Quant. Grav. {\bf 23}, S51 (2006).
\bibitem[2]{s4burst}
B.~Abbott \emph{et al}, Class. Quant. Grav. {\bf 24}, 5343 (2007).
\bibitem[3]{s1burst}
B.~Abbott \emph{et al}, Phys. Rev. D {\bf 69}, 102001 (2004).
\bibitem[4]{s2burst}
B.~Abbott \emph{et al}, Phys. Rev. D {\bf 72}, 062001 (2005).
\bibitem[5]{s3burst}
B.~Abbott \emph{et al}, Class. Quant. Grav. {\bf 23}, S29 (2006).
\bibitem[6]{s4burstgeo}
B.~Abbott \emph{et al}, Class. Quant. Grav. {\bf 25} 245008 (2008).
\bibitem[7]{OttBurrows}
C.D.~Ott, Class. Quant. Grav. {\bf 26} 063001 (2009).
\bibitem[8]{brwave1}
L.~Baiotti and L. ~Rezzolla, Phys. Rev. Lett. {\bf 97}, 141101 (2006).
\bibitem[9]{brwave2}
L.~Baiotti \emph{et al}, Class. Quant. Grav. {\bf 24}, S187 (2007).
\bibitem[10]{HMNS}
R.~Oechslin and H.-T.~Janka, Phys. Rev. Lett. {\bf 99}, 121102 (2007).
\bibitem[11]{HMNS2}
K.~Kiuchi \emph{et al}, submitted to Phys. Rev. D. arXiv:0904.4551 [gr-qc]
\bibitem[12]{fmodes}
B.F.~Schutz, Class. Quant. Grav. {\bf 16}, A131 (1999).
\bibitem[13]{nsm}
J.G.~Jernigan, AIP Conf. Proc. {\bf 586}, 805 (2001).
\bibitem[14]{lmbh}
K.T.~Inoue and T.~Tanaka, Phys.Rev.Lett. {\bf 91}, 021101 (2003).
\bibitem[15]{SGR}
J.E. Horvath, Modern Physics Lett. A {\bf 20}, 2799 (2005).
\bibitem[16]{cosmicstring}
H.J.~Mosquera Cuesta and D.M.~Gonzalez, Phys. Lett. B {\bf 500}, 215-221 (2001).
\bibitem[17]{virgo}
F.~Acernese \emph{et al} Class. Quant. Grav. {\bf 23}, S63 (2006).
\bibitem[18]{shourovthesis}
S.~Chatterji, MIT Ph.D. thesis, September 2005.
\bibitem[19]{qtransform}
S.~Chatterji \emph{et al}, Class. Quant. Grav. {\bf 21}, S1809 (2004).
\bibitem[20]{lowfreq}
B.~Abbott \emph{et al} (LSC), submitted to Phys. Rev. D. gr-qc/0905.0020 
\bibitem[21] {ht calibration}
X.~Siemens \emph{et al}, Class. Quant. Grav. {\bf 21}, S1723 (2004).
\bibitem[22]{lpf2}
J.~Makhoul, Proc. IEEE {\bf 63} (1975).
\bibitem[23]{CorrPower}
L.~Cadonati and S.~M$\acute{\rm{a}}$rka, Class. Quant. Grav. {\bf 22}, S1159 (2005).
\bibitem[24]{cwb}
S.~Klimenko \emph{et al}, Class. Quant Grav. {\bf 25}, 114029 (2008).
\bibitem[25]{xpipeline}
S.~Chatterji \emph{et al}, Phys. Rev. D {\bf 74}, 082005 (2006).
\bibitem[26]{kleinewelle}
S.~Chatterji, L.~Blackburn, G.~Martin, and E.~Katsavounidis, Class. Quant. Grav. {\bf 21}, S1809-18 (2004).
\bibitem[27]{highfreqantpat}
M.~Rakhmanov, J.D.~Romano and J.T.~Whelan, Class. Quant. Grav. {\bf25}, 184017 (2008).
\bibitem[28]{GravEn}
A.L.~Stuver and L.S.~Finn, Class. Quant. Grav. {\bf 23}, S799 (2006).
\bibitem[29]{timelist}
https://dcc.ligo.org/cgi-bin/private/DocDB/ShowDocument?docid=5982
\bibitem[30]{IGEC}
P. ~Astone \emph{et al}, Phys. Rev. D {\bf 76}, 102001 (2007).
\bibitem[31]{enhanced}
J.R.~Smith \emph{et al}, submitted to Class. Quant. Grav. gr-qc/0902.0381v2
\bibitem[32]{advancedligo}
P.~Fritschel \emph{Gravitational-Wave Detection: Proc. SPIE} (2003) vol 4856 ed M.~Cruise and P.~Saulson (Bellingham, WA:SPIE Optical Engineering Press) pp 282-91.
\bibitem[33]{advancedvirgo}
F.~Acernese \emph{et al} Class. Quant. Grav. {\bf 23}, S635 (2006).


\end{thebibliography}
\end{document}